\documentclass[lettersize,journal]{IEEEtran}

\usepackage{amsmath,amsfonts}
\usepackage{array}
\usepackage{subfigure}
\usepackage{textcomp}
\usepackage{stfloats}
\usepackage{url}
\usepackage{verbatim}
\usepackage{graphicx}
\def\BibTeX{{\rm B\kern-.05em{\sc i\kern-.025em b}\kern-.08em
    T\kern-.1667em\lower.7ex\hbox{E}\kern-.125emX}}
\usepackage{balance}

\usepackage[justification=centering]{caption}

\usepackage{booktabs,multirow}
\usepackage{placeins}
\usepackage{algpseudocode,algorithm}
\usepackage{color}
\usepackage[table,xcdraw]{xcolor}
\usepackage{enumerate}   
\usepackage{adjustbox}
\usepackage[normalem]{ulem}
\usepackage{amssymb}             
\usepackage{tikz,hyperref}
\definecolor{lime}{HTML}{A6CE39}
\DeclareRobustCommand{\orcidicon}{%
	\begin{tikzpicture}
	\draw[lime, fill=lime] (0,0) 
	circle [radius=0.16] 
	node[white] {{\fontfamily{qag}\selectfont \tiny ID}};	\draw[white, fill=white] (-0.0625,0.095) 
	circle [radius=0.007];	\end{tikzpicture}
	\hspace{-2mm}}
\foreach \x in {A, ..., Z}{%
	\expandafter\xdef\csname orcid\x\endcsname{\noexpand\href{https://orcid.org/\csname orcidauthor\x\endcsname}{\noexpand\orcidicon}}
	}

\hypersetup{hidelinks}

\usepackage[capitalise]{cleveref}

\newcommand{\vct}[1]{\boldsymbol{\mathbf{#1}}} 
\newcommand{\mat}[1]{\mathbf{#1}} 

\begin{document}

\title{Attention and DCT based  Global Context Modeling for Text-independent Speaker Recognition}

\author{Wei Xia, \textit{Student Member, IEEE},
        and~John~H.~L.~Hansen, \textit{Fellow, IEEE}
\thanks{Initial manuscript completed on March 10, 2022; accepted to IEEE/ACM Transactions on Audio, Speech, and Language Processing. 
This work was supported by the University of Texas at Dallas from the Distinguished University Chair
in Telecommunications Engineering held by John H. L. Hansen.}%
\thanks{The authors are with Center for Robust Speech Systems, Erik Jonsson School of Engineering University of Texas at Dallas, Richardson, TX 75080, USA (e-mail: wei.xia@utdallas.edu; john.hansen@utdallas.edu).}%
}

%

\maketitle

\begin{abstract}

Learning an effective speaker representation is crucial for achieving reliable performance in speaker verification tasks.
Speech signals are high-dimensional, long, and variable-length sequences that entail a complex hierarchical structure. Signals may contain diverse information at each time-frequency (TF) location.
The standard convolutional layer that operates on neighboring local regions often fails to capture the complex TF global information. 
Our motivation stems from the need to alleviate these challenges by increasing the modeling capacity, emphasizing significant information, and suppressing possible redundancies in the speaker representation. We aim to design a more robust and efficient speaker recognition system by incorporating the benefits of  attention mechanisms and Discrete Cosine Transform (DCT) based signal processing techniques, to effectively represent the global information in speech signals.
To achieve this, we propose a general global time-frequency context modeling block for speaker modeling.
First, an attention-based context model is introduced to capture the long-range and non-local relationship across different time-frequency locations. Second, a 2D-DCT based context model is proposed to improve model efficiency and examine the benefits of signal modeling. A multi-DCT attention mechanism is presented to improve modeling power with alternate DCT base forms.
Finally, the global context information is used to recalibrate salient time-frequency locations by computing the similarity between the global context and local features.
The proposed lightweight blocks can be easily incorporated into a speaker model with little additional computational costs. This effectively improves the speaker verification performance compared to the standard ResNet model and Squeeze\&Excitation block by a large margin.
Detailed ablation studies are also performed to analyze various factors that may impact performance of the proposed individual modules.
Our experimental results show that the proposed global context modeling method can efficiently improve the learned speaker representations by achieving channel-wise and time-frequency feature recalibration.
\end{abstract}


%
\maketitle

\section{Introduction}
\label{sec:intro}

\IEEEPARstart{A}{utomatic} Speaker Verification (ASV) involves determining a person's identity from an input audio stream. ASV provides a natural and efficient way for biometric identity authentication. 
The ability to perform speaker verification is helpful in retrieving target individuals in many practical applications. Speaker recognition can be used for audio surveillance~\cite{foggia2016audio}, computer access control, and voice authentication for telephony scenarios~\cite{lee2011joint,crocco2016audio}.
It is also helpful for targeted speaker extraction systems if we have good speaker embeddings~\cite{wang2018voicefilter,xie2019multi,chen2021scenario,medennikov2020target}.  When multi-speakers in a meeting, multi-talker speaker tracking~\cite{yu2017permutation} is beneficial for analyzing each person's opinion, emotion, and engagement. Smart home devices, including Google Home, Amazon Alexa, and Apple Homepod, could also benefit from ASV for personalized voice applications~\cite{lau2018alexa, rahman2018attention}.

Learning an effective speaker representation is crucial in speaker verification tasks. The paradigm has shifted from GMM-UBM and factor analysis based methods include i-vector~\cite{matvejka2011full,hansen2015speaker} with a probabilistic linear discriminant (PLDA) back-end~\cite{kenny2010bayesian,prince2007probabilistic} towards deep neural network based models. Neural network architectures such as ResNet~\cite{he2016deep} and Time-Delay Neural Network~\cite{snyder2018x,snyder2019speaker,zhang2021improving,mohammadamini2022learning} and Res2Net~\cite{zhou2021resnext,desplanques2020ecapa,hong2022generalization} have been explored to improve the speaker embedding extraction. 
Margin based softmax loss functions such as Angular Softmax~\cite{liu2017sphereface}, Additive Margin Softmax~\cite{wang2018additive}, Additive Angular Margin~\cite{deng2019arcface}, Adaptive Margin~\cite{wang2020adaptive}, and Triplet loss~\cite{zhang2018text}, have been shown to be effective in learning discriminative speaker embeddings. 
Several new temporal pooling methods, including attentive pooling~\cite{okabe2018attentive,india2019self}, Spatial Pyramid Pooling~\cite{jung2019spatial}, and LDE~\cite{cai2018exploring} can aggregate variable length input features to a fixed-length utterance level representation. 
Various noise and language robust speaker recognition models~\cite{Vinay2022SkipConvGAN,xia2019cross,mamun2019quantifying,joglekar2019fearless}, training paradigms~\cite{jung2019rawnet,Heo:2017ci}, and domain adaptation~\cite{wang2020cross,wang2021multi} methods have been proposed and significantly improve speaker verification system performance. 
 
One basic building block for most SV models is the convolutional layer, which learns filters to capture local patterns. However, the filter that only operates on the neighboring local context often fail to capture long-range, non-local global information. Our motivation of this study is to design a mechanism to effectively aggregate information among different TF units, reduce channel redundancies, and strengthen significant time-frequency locations by channel recalibration and time-frequency enhancement. Speech signals contain a rich and diverse array of information at each TF location, which plays a vital role in accurately identifying speakers. Focusing on significant TF locations, such as salient regions in the spectrogram, could yield better speaker representations. This necessitates the development of an effective mechanism to aggregate and leverage information from different TF units.

The attention mechanism is one way to aggregate information at different time-frequency locations.
Many recent studies~\cite{hu2018squeeze,bello2019attention,cao2019gcnet,zhang2021duality,xia2020speaker} attempt to alleviate these issues by improving the encoding of TF information. 
One popular approach to accomplish this is the ``Squeeze \& Excitation'' (SE) block~\cite{hu2018squeeze,xia2019sound}, which explicitly models the inter-dependencies between channels of the feature map. This SE block factors out the time-frequency dependency by average pooling to learn a channel specific descriptor, which is used to rescale the input feature map to highlight only salient channels. CBAM~\cite{woo2018cbam,yadav2020frequency} uses both global average pooling and global max pooling to obtain better performance.
Non-local network~\cite{wang2018non} computes the response at a position as a weighted sum of the features at all positions to capture long-range dependencies. ECA-Net~\cite{wangeca} that uses cross-channel interaction can preserve performance while significantly decreasing model complexity. SkNet and ResNeSt~\cite{li2019selective,zhang2020resnest} introduce a split attention strategy with multiple branches along with different kernel sizes. The information in these branches is fused to obtain a richer representation. 

Discrete Cosine Transform (DCT) is another way to effectively aggregate meaningful TF information. From a signal processing perspective, another motivation for this study is the efficient utilization of Discrete Cosine Transform (DCT) in the context of speaker verification tasks. Leveraging the signal compression properties of DCT allows our model to focus on the most relevant information while discarding potential redundancies present in the audio signal. By incorporating DCT-based models in speaker verification tasks, we can effectively exploit the benefits of signal compression and frequency domain learning for more accurate and efficient speaker recognition systems. DCT has been widely used for audio and image compression~\cite{Wang_2020_CVPR}. For example, Wu et al.~\cite{zhang2020davd} formulated a DCT based deep network for video action recognition that used a separate network for i-frames and p-frames. Ghosh et al.~\cite{ghosh2016deep} used DCT as part of their network’s first layer and showed that it speeded up convergence for training. Ulicny et al.~\cite{ulicny2017using} created separate filters for each DCT basis function. DCT has been shown to be an efficient tool for frequency domain learning~\cite{qin2021fcanet,ehrlich2019deep} for image classification.
This combination of deep learning architectures and signal processing techniques has the potential to  advance speaker recognition models and broaden the applicability of these systems in different scenarios.

A critical step in this modeling phase is to efficiently encode the feature map with a vector to preserve the collective information as much as possible. Global Average Pooling (GAP) is the most common choice to encode the time-frequency information at a very small computational cost. However, GAP may cause inferior results in some cases and may not capture comprehensive context information across distinct time and frequency locations. SOCA~\cite{dai2019second} proposes a second-order channel attention module to adaptively rescale the channel-wise features by using second-order feature statistics for more discriminative representations.

In this study, first, we apply a learnable attentive time-frequency context embedding to efficiently model the global speech contextual information.
We show that GAP is actually a special case of attention, and 2D-DCT based context models (lowest frequency basis).
When GAP is used to embed the time-frequency information into a scalar, it is possible to lose some rich audio information. 
A query independent attention is applied to emphasize important time-frequency locations.
Second, we propose a 2D-DCT method to embed the time-frequency information using different signal bases.
With the pre-computed 2D-DCT weights, we design a multi-DCT attention method to leverage more frequency components for global context modeling. 
Last, after the channel wise recalibration using the attention/DCT based global context model, we propose a time-frequency enhancement step by leveraging the global information to emphasize the important local time-frequency vector bins at different positions.
The whole procedure can be summarized as a global time-frequency context modeling approach built upon the Squeeze\&Excitation block.
Experimental results show that with the proposed methods, the Equal Error Rate (EER) and the minimum Detection Cost Function (DCF) are reduced significantly.

The main contributions of this work are summarized as follows.
\begin{itemize}
\item A generalized global context modeling (GCM) method is introduced by emphasizing important channel and time-frequency regions of speaker representations.
\item A query-independent attention based global context model (Att-GCM) is proposed to improve the long-range and non-local relationship across different time-frequency locations.
\item A 2D-DCT based global context model (DCT-GCM) is also proposed to improve model efficiency and examine the benefits of signal modeling. A multi-DCT attention mechanism is introduced to improve the modeling power with different DCT bases.
\item A time-frequency enhancement (TFE) method is proposed to leverage the global context vector as an intermediate representation. This can further enhance the global context models by performing time-frequency wise recalibration.
\item Finally, extensive experiments are performed to demonstrate the effectiveness of our proposed methods in both accuracy and computational complexity.
\end{itemize}

\vspace{2ex}
In the following sections, we review the squeeze and excitation method in Sec. \ref{sec:se}. Global time-frequency context modeling method and the attention based context model are described in Sec. \ref{sec:gtfc}. Sec. \ref{sec:dct} describes the 2D-DCT based context model. The time-frequency enhancement method is explained in Sec. \ref{sec:tfe}. We provide detailed formulation and explanations of our experiments in Sec. \ref{sec:exp}, as well as results and discussions in Sec. \ref{sec:result}. Finally, we conclude our work in Sec. \ref{sec:conclusion}.
In Table I, we provide a list of abbreviations for a better readability and comprehension of the approaches discussed in this paper.
\begin{table}[htbp]
  \centering
  \caption{List of abbreviations and definitions used in the paper.}
  \begin{adjustbox}{max width=1\linewidth}
    \setlength{\tabcolsep}{10pt}
  \begin{tabular}{ll}
    \toprule
    Abbreviation & \multicolumn{1}{c}{Definition}  \\
    \hline
    SE & Squeeze and Excitation   \\
    Att-GCM   & Attention based Global Context Model  \\
    DCT-GCM & DCT based Global Context Model  \\
    TFE   & Time-frequency Enhancement  \\
    DCT   & Discrete Cosine Transform  \\
    GTFC   & Global Time-frequency Context  \\
    \bottomrule
    \end{tabular}
  \end{adjustbox}
\label{tab:abbr}
\end{table}

\begin{figure}[bp]
    \centering
    \includegraphics[width=0.93\linewidth, height=8cm]{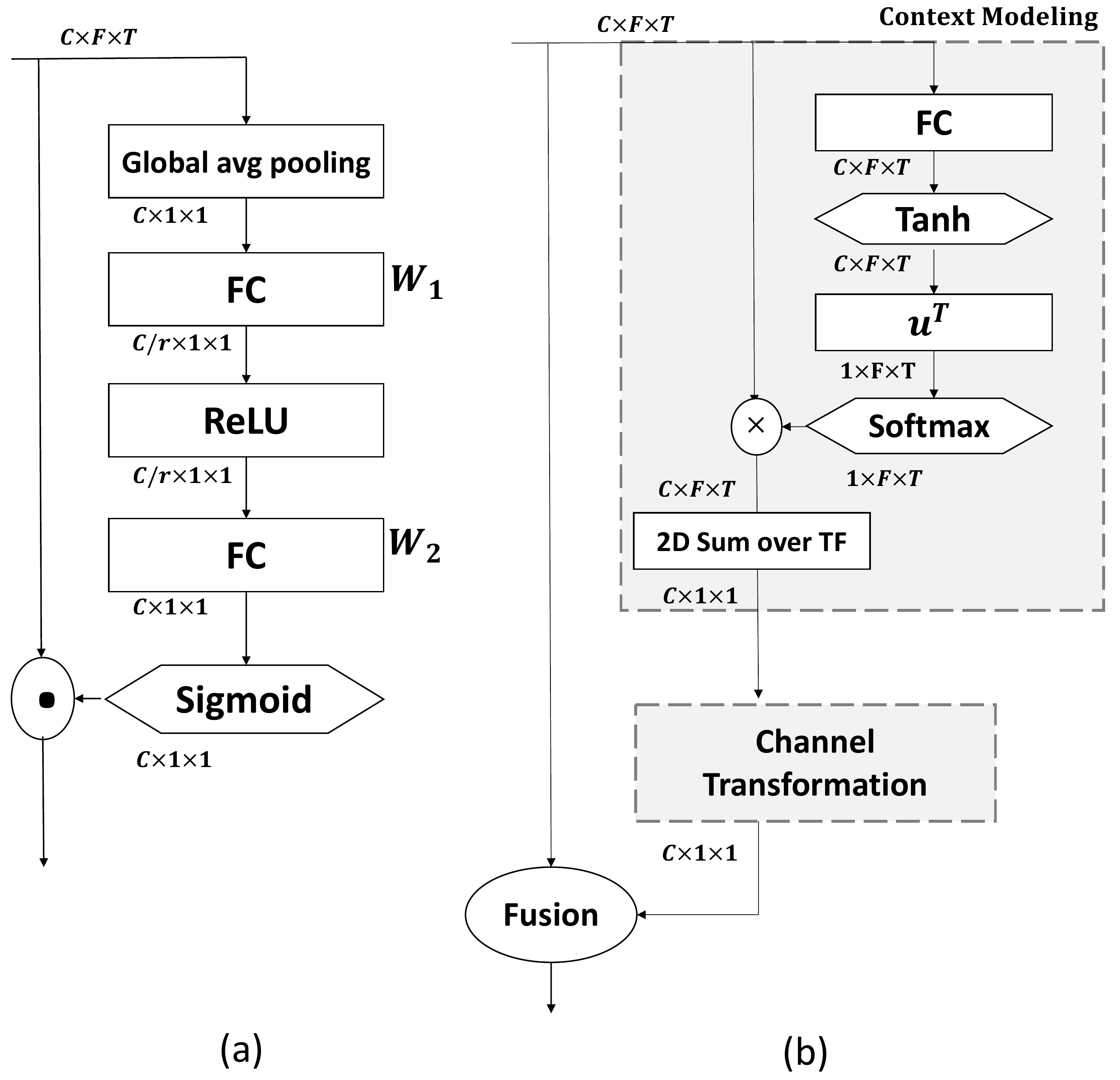}
    \caption{(a) SE block. (b) Proposed Attention based global time-frequency context modeling method and channel-wise transformation.}
    \label{fig:se1}
\end{figure}

\vspace{1ex}
\section{Recap of Squeeze and Excitation attention}
\label{sec:se}

The Squeeze and Excitation (SE) channel attention uses a squeeze operation to summarize time-frequency information into a channel embedding. This block is illustrated Fig. \ref{fig:se1}(a). We define the feature map of the input audio as $\mathbf{X}=\left[\mathbf{X}_{1}, \mathbf{X}_{2}, \ldots, \mathbf{X}_{c}\right]$, where $\mathbf{X}_{c} \in \mathbb{R}^{F \times T}$ is the feature matrix of channel $c$. In this procedure, a global average pooling layer is applied first to generate a channel-wise embedding $\mathbf{g} \in \mathbb{R}^{C}$ with its $c$-th element,
\begin{align}
g_{c}=\mathcal{F}_{s q}\left(\mathbf{X}_{c}\right)=\frac{1}{F \times T} \sum_{f=0}^{F-1} \sum_{t=0}^{T-1} \mathbf{X}_{c}(f, t)
\end{align}
This operation embeds the global time-frequency information into the intermediate output vector $\mathbf{g}$. This vector contains the statistics expressing the input signal's collective time-frequency content. In order to capture the channel-wise dependencies, a gating mechanism with a sigmoid activation function is used to learn a nonlinear relationship between channels as follows,
\begin{align}
\mathbf{s}=\mathcal{F}_{e x}(\mathbf{g}, \mathbf{W})=sigmoid\left(\mathbf{W}_{2} ReLU\left(\mathbf{W}_{1}  \mathbf{g}\right)\right)
\end{align}

\noindent where $\mathbf{W}_{1} \in \mathbb{R}^{\frac{C}{r} \times C}$ and $\mathbf{W}_{2} \in \mathbb{R}^{C \times \frac{C}{r}}$ are the weights of two fully-connected layers. The dimensionality reduction factor $r$ indicates the bottleneck in the channel excitation block. Note that the original channel dimension is recovered by the second $\mathrm{FC}$ layer. With a sigmoid activation function, the channel-wise attention vector $\mathbf{s} \in \mathbb{R}^{C}$ is obtained. Finally, $U$ is recalibrated with the attention vector as,
\begin{align}
\hat{\mathbf{X}}_{c}=\mathcal{F}_{\text {scale }}\left(\mathbf{X}_{c}, s_{c}\right)=s_{c} \mathbf{X}_{c}
\end{align}

\noindent Here, $\hat{\mathbf{X}}=\left[\hat{\mathbf{X}}_{1}, \hat{\mathbf{X}}_{2}, \ldots, \hat{\mathbf{X}}_{C}\right]$ is the final set of channel-wise recalibrated features. $s_{c}$ is the $c$-th element of the vector $\mathbf{s}$. In this block, the input features are attentively scaled so that the important channels are emphasized, and the less important ones are diminished.

\section{Attention based context modeling}
\label{sec:gtfc}

The global context information and channel relationship in the SE-Net are inherently implicit. The global average operation treats each time-frequency location the same and therefore does not consider the relationship or rich information of each time-frequency bin.
To better model long-range relationships and local interactions,
a generalized method is proposed based  on the global context modeling for channel and time-frequency wise feature recalibration. A query-independent attention map is computed. For each time-frequency location, this attention mechanism is used to learn the weight and subsequently pool the corresponding feature values to obtain a global representation.

To summarize, an attention mechanism is first used to learn a global time-frequency embedding $\vct g \in \mathbb{R}^{C}$. Next, a channel-wise transformation is applied by broadcasting the global TF context vector to each channel.
In Fig. \ref{fig:se1}(b), the process is shown on how the Global Time-Frequency Context (GTFC) embedding is learnt and applied for the channel-wise feature map enhancement: (a) the context modeling module groups the features of all positions together via global attentive pooling; 
(b) local channel interactions are applied next on the GTFC to capture channel-wise dependencies; 
(c) a fusion function is used to distribute the context vector across channels. In the following sections, further details about this process are described in detail.

\subsection{Global Attention Pooling}

Proposed in our prior work \cite{xia2018speaker}, a global context embedding module is designed to aggregate the non-local, long-range time-frequency relationship in each channel. Since individual TF locations may reflect a range of content importance for SV, an attention mechanism is used to focus on salient regions that may have a more significant impact on the global context.
The module can exploit comprehensive contextual information outside the small receptive fields of the convolutional layers to better encode global TF information. Given an input feature map $\mat{X_c} \in \mathbb{R}^{F \times T} $, the following module is designed,
\begin{align}
g_{c} = &  \sum_{f=0}^{F-1} \sum_{t=0}^{T-1} \alpha_{f,t} \mat{X}_{c}(f,t)
\end{align}
\noindent where $\alpha_{f,t}$ is the learned attention weight at a time-frequency location $(f,t)$. It is from the attention scalar score $e_{f,t}$, which is computed through an MLP $\mat{W}_{\alpha}$ on the feature vector $\vct{x}_{f,t} \in \mathbb{R}^{C}$ and a hidden vector $\vct {u}_{\alpha}$,
\begin{align}
e_{f,t} & = \vct{u}_{\alpha}^{\mathsf{T}} tanh(\mat{W}_{\alpha} \vct{x}_{f,t} + \vct{b} ) + k
\end{align}
where $\vct{b}$ and $k$ are bias terms. The attention score is normalized over all time-frequency locations by a softmax function so as to sum to the unity,
\begin{align}
\alpha_{f,t} & = \frac{exp(e_{f,t})}{\sum\limits_{f=0}^{F-1} \sum\limits_{t=0}^{T-1} exp(e_{f,t})}
\end{align}

The attention unit is efficient in representing the nonlinear, complex activations, and is a general form of mean or max pooling. The squeeze-excitation (SE) block is also an instantiation of our proposed method, where $\alpha_{f,t} = \frac{1}{F\times T}$.
Additionally, each time-frequency location shares one attention mask. Therefore, this query-independent attention mechanism helps efficiently learn the weight at each location to achieve a better global context representation. 
Since this proposed context modeling block is lightweight, it can be applied in multiple layers to capture the long-range dependency with only a slight increase in computation cost. Considering ResNet-34 as an example, ResNet-34 + Att-GCM denotes the addition of the proposed attention based global context modeling block to the last layer in ResNet-34 with a bottleneck ratio of 8. This proposed block only increases the overall system with 0.40M additional parameters.

\subsection{Local Cross Channel Interaction}
\label{sec:local_channel}
After obtaining the time-frequency context vector, it would be useful to compare two alternative ways to perform local channel interactions to capture the channel relationships.

\begin{itemize}
    \item \textbf{Fully connected layer.} Same as the SE module, two FC layers are applied with dimension reduction $r$ to reduce the model parameters. This module introduces $2\times C^2/r$ additional parameters, where $C$ is the number of channels.

    \item \textbf{1D Convolution}. It is also possible to use 1D convolution with a kernel size of $k$ to perform channel interactions. This operation only introduces $k$ additional parameters. In \cite{Wang_2020_CVPR}, the 1D convolution operation is introduced as an Efficient Channel Attention (ECA), which is followed by an adaptive kernel size selection method, i.e., $k = \phi(C) = |\frac{log_2(C)}{\gamma} + \frac{b}{\gamma}|$. Here, we set $\gamma = 2$ and $b = 1$ and compare with the FC layers in subsequent experiments.
\end{itemize}

Finally, we aggregate the transformed global context vector to recalibrate the feature map across channels with a sigmoid function and element-wise dot product operation.

\begin{figure*}[ht]
\centering
\subfigure[Proposed multi-DCT attention]{\includegraphics[width=0.85\linewidth, height=5.5cm]{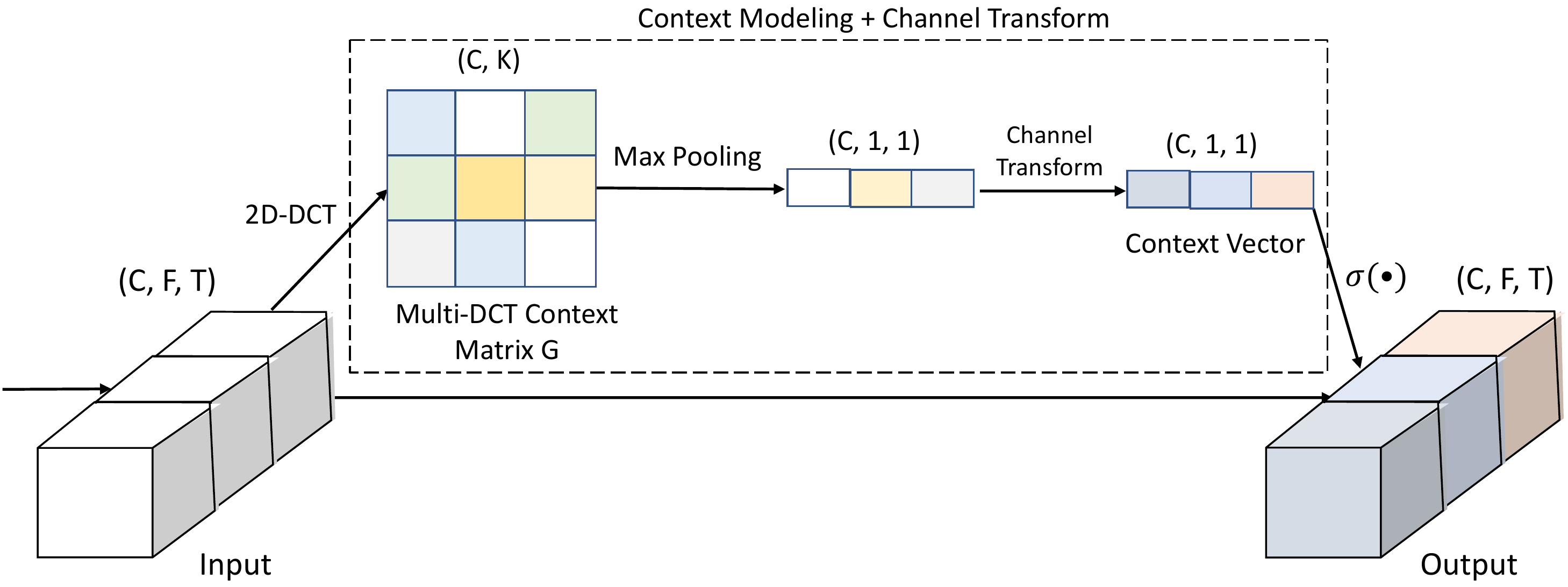}}
\subfigure[An example of $8 \times 8$ DCT bases]{\includegraphics[width=0.45\linewidth, height=5.5cm]{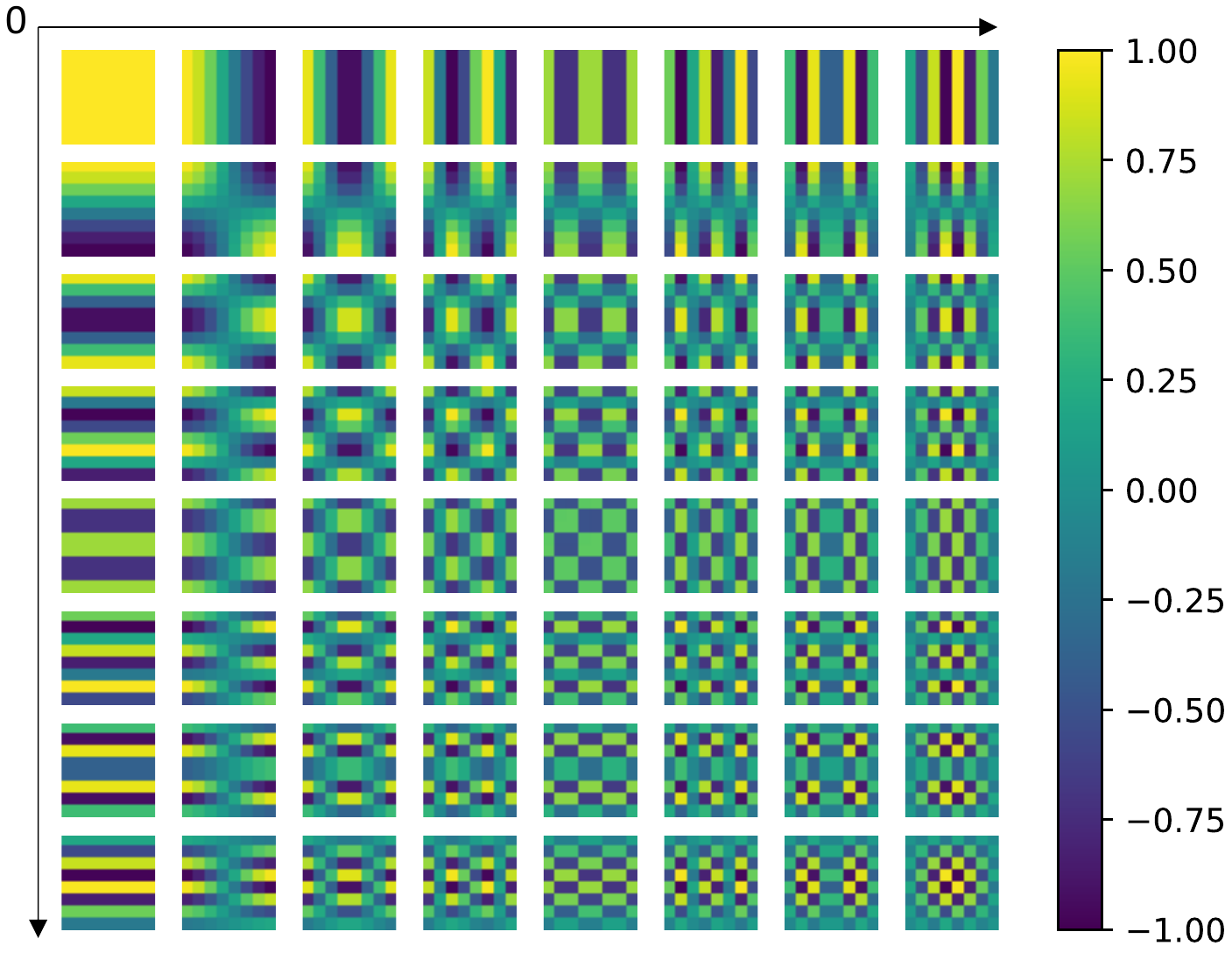}}
\caption{Pipeline of the proposed two-dimensional Discrete Cosine Transform based time-frequency context model.}
\label{fig:dct} 
\end{figure*}

\section{DCT based context modeling}
\label{sec:dct}
In the previous section, an attention mechanism was proposed to learn the weights of each time-frequency location of the feature map. The learned weights are parametric and purely data-driven. This section further introduces a data-independent and explainable method to compute the weights of each time-frequency location. We apply the two dimensional Discrete Cosine Transform (2D-DCT) based pooling to represent rich global context information. We also show that this method is a generalized form of Global Average Pooling (GAP).

\subsection{Discrete Cosine Transform}

Given an input feature map $\mat{X}_c \in \mathbb{R}^{F\times T}$, its 2D Discrete Cosine Transform $g$ is formulated as,

\begin{align}
g_{c,i,j} &= \textrm{2D-DCT}\left(\mat{X}_{c}\right) \nonumber \\
&=\sum_{f=0}^{F-1} \sum_{t=0}^{T-1}  B_{i, j}^{f, t} \mat{X}_c(f,t)
\end{align}

\noindent where index $i \in\{0,1, \cdots, F-1\}, j \in\{0,1, \cdots, T-1\}$, and $B_{i,j}$ is a basis function of the 2D-DCT. It is written as,
\begin{align}
B_{i, j}^{f, t}=\cos \left(\frac{\pi i}{F}\left(f+\frac{1}{2}\right)\right) \cos \left(\frac{\pi j}{T}\left(t+\frac{1}{2}\right)\right)
\end{align}

The 2D-DCT based pooling is applied to aggregate the input feature map. The weights correspond to a specific DCT basis. For example, we visualize 64 DCT bases in Fig. \ref{fig:dct}(b). It can be proven that GAP used in SE channel attention is a special case of 2D-DCT, where $i$ and $j$ are set to 0. In this case, this can be written as,
\begin{align}
g_{c,0,0} &=\sum_{f=0}^{F-1} \sum_{t=0}^{T-1}  \cos \left(\frac{0}{F}\left(f+\frac{1}{2}\right)\right) \cos \left(\frac{0}{T}\left(t+\frac{1}{2}\right)\right) \mat{X}_c(f, t) \nonumber \\ 
&=\sum_{f=0}^{F-1} \sum_{t=0}^{T-1} \mat{X}_c(f, t) 
\end{align}
\noindent where $g_{c,0,0}$ represents the lowest frequency component of 2D-DCT in the c-th channel, and it is proportional to GAP.

\subsection{Multi-DCT Channel Attention}
\label{sec:multi_dct}
The previous section has shown that GAP in the SE channel attention is the lowest frequency component of 2D-DCT. In order to use rich speech information for the global context modeling, the following mechanism is designed to leverage multiple DCT components and select the maximum response for each channel. The entire process is illustrated in Fig. \ref{fig:dct}(a).

Here, denote $\mathbf{X}=\left[\mathbf{X}_{1}, \mathbf{X}_{2}, \ldots, \mathbf{X}_{c}\right]$, where $\mathbf{X}_{c} \in \mathbb{R}^{F \times T}$ is the feature matrix of channel $c$. 
For one DCT component $B_{i, j}$, where $\left[i, j\right]$ are indexes of a DCT component, we can compute a 2D-DCT transform of $\mathbf{X}_{c}$ use equation (7). With the $K$ DCT components, we can obtain a $K$ dimensional vector $\vct{\phi}_c$,

\begin{align}
\vct{\phi}_c &= \begin{bmatrix}
           \textrm{2D-DCT}^0(\mat{X}_{c}) \\
           \textrm{2D-DCT}^1(\mat{X}_{c}) \\
           \vdots \\
           \textrm{2D-DCT}^{K-1}(\mat{X}_{c})
         \end{bmatrix}
\end{align}

The complete 2D-DCT global context embedding matrix can be defined as $\mat{G} = [\vct{\phi}_0^{\mathsf{T}}; \vct{\phi}_1^{\mathsf{T}}; \cdots; \vct{\phi}_c^{\mathsf{T}}]$, where $\mat{G} \in \mathbb{R}^{C\times K} $. Next, a row-wise max operation is applied to select the maximum response for each channel to obtain the final global context vector $\vct{g} \in \mathbb{R}^{C}$.
\begin{align}
\vct{g} = \underset{k}{\max} \ \mat{G}(:, k)
\end{align}
\noindent We apply the same local channel interaction and re-calibration steps discussed previously in section \ref{sec:local_channel} to adjust the channel-wise feature map.

\noindent \textbf{DCT component selection.} Another important issue in this method is the selection process of the $K$ DCT components. It is impractical and problematic to choose the best $K$ DCT components with the highest performance on the test data since it requires an exhaustive set of experimental results. Essentially it is impractical to run experiments for every DCT component to create a ranked order. The ranking order also changes with different datasets. 
In our experiments, DCT components are pre-determined and only related to specific time and frequency indices. For instance, in Fig. \ref{fig:dct}(b), we show an example of 64 DCT components.  
They are ordered from (0,0) (top left), (0,1), (1,0), (0,2), (1,1)..., up to (8,8) (bottom right) from the lowest to the highest component. We only select the lowest $K$ components.  
They are ordered by the sum of 2D indexes (i+j). If two sums are the same, the one with a smaller index i has the lower order. We only select the lowest $K$ components. 

If the input feature size is fixed, we can divide the DCT space according to the smallest feature map size. If input data has a variable length, we can chunk the data or use adaptive average pooling to rescale the feature map to a fixed size.

\begin{figure*}[htbp]
  \centering
  \includegraphics[width=0.8\linewidth, height=6.5cm]{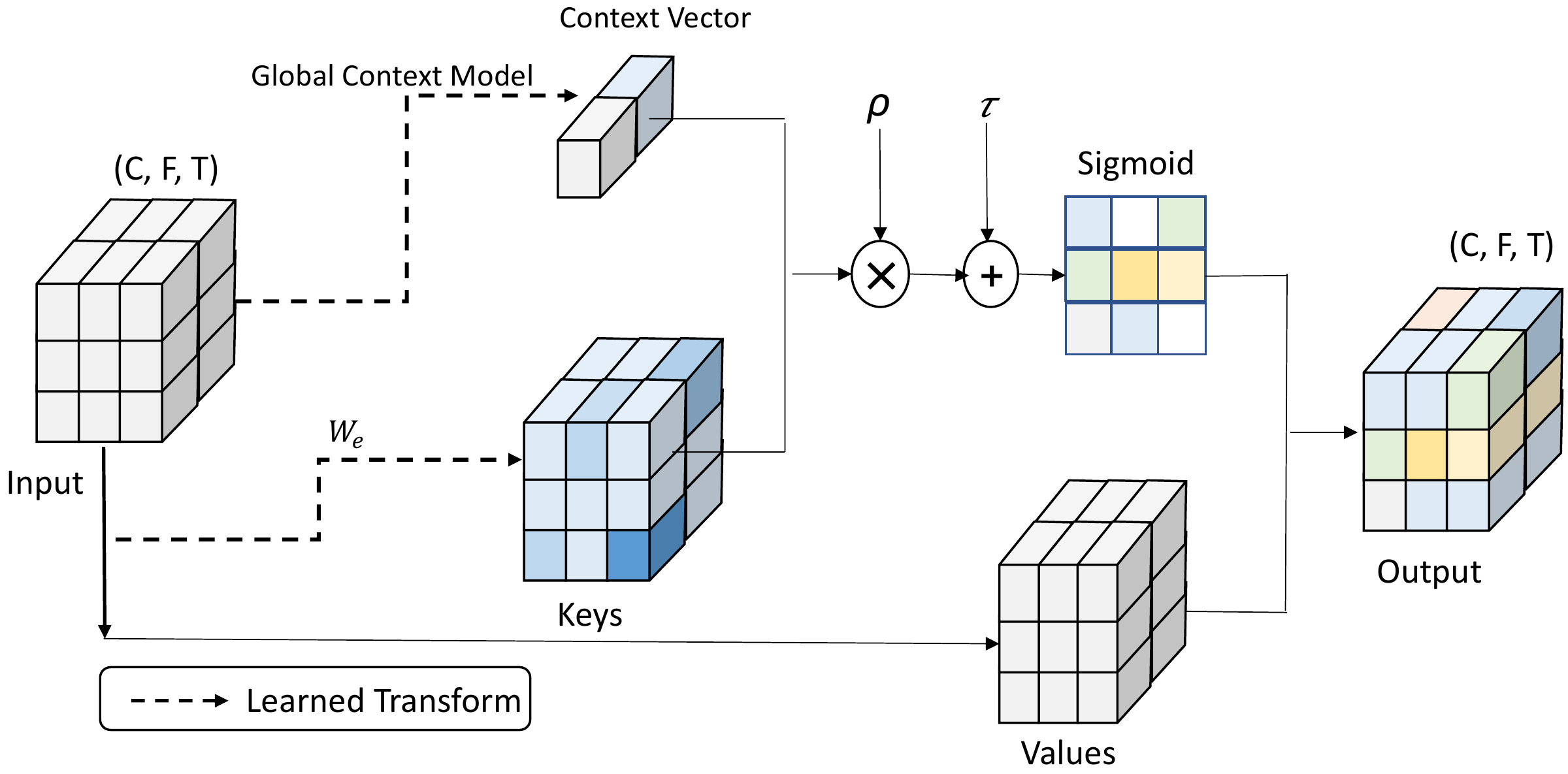}
  \caption{Time-frequency feature recalibration using a group-wise Time-frequency Enhancement (TFE) between the GTFC embedding and the local TF feature vector. The TFE module is applied after the channel recalibration.}
  \label{fig:se2}
\end{figure*}

\section{Time-frequency enhancement}
\label{sec:tfe}
We also propose a method for computing the TF attention map based on the correlation between the global TF context (GTFC) embedding and the local feature vectors. After the channel recalibration, this method is applied to further improve the proposed context models by strengthening significant time-frequency locations.
In Fig.~\ref{fig:se2}, we illustrate the proposed group wise time-frequency enhancement (TFE) method.
First, the $C$ channels, $F\times T$ convolutional feature map is divided into $N$ groups along the channel dimension. It is assumed that each group could gradually learn a specific response during the training process. In each group, we have a set of local feature vectors $ \mathcal X =\{\vct{x}_{f,t}\}, \vct{x}_{f,t} \in \mathbb{R}^{C/N}, f \in \{0,1, \cdots, F-1\}, t \in\{0,1, \cdots, T-1\} $.
Ideally, it would be good to obtain features with strong responses at salient time-frequency positions (e.g., high energy region). 
However, both disturbances and reverberations may prevent us from obtaining the appropriate neuron activations following convolution.
To alleviate this problem, the group-wise normalized GTFC embedding $\hat{\vct{g}}$ is used as a global group representation, and the correlation is computed between the GTFC vector and the local feature vector $\vct{x}_{f,t}$ at each time-frequency location. The similarity score is calculated as follows,
\begin{align}
  e_{f,t} & = \text{score}(\hat{\vct{g}}, \vct{x}_{f,t}) = \hat{\vct{g}}^{\mathsf{T}} \mat{W}_e \vct{x}_{f,t}
\end{align}

\noindent For each group, we compute a corresponding importance coefficient $e_{f,t}$ at each position, using the general dot product scoring function~\cite{luong2015effective} in Eq. (12). 
$\mat{W}_e$ is a weight matrix to be learned. Additionally, $\vct{e}$ needs to be normalized over the time-frequency domain to prevent the biased magnitude of coefficients between various samples,
\begin{align}
  \hat{e}_{f,t} = \frac{e_{f,t} - \mu_e}{\sigma_e + \epsilon}, \mu_e = \frac{\sum_{f,t}e_{f,t}}{F\times T}, \sigma_e^2 = \frac{\sum_{f,t}^{}(e_{f,t} - \mu_e)^2}{F\times T}
\end{align}
\noindent where $\epsilon$ (e.g., $1e^{-5}$) is a constant for numerical stability. We provide a pair of parameters ($\rho, \tau$) for each coefficient $e_{f,t}$ to ensure that the normalization introduced in the network can represent the identity transformation. The parameters scale and shift the normalized value. Finally, to obtain the enhanced feature vector $\hat{\vct{x}}_{f,t}$, the original $\vct{x}_{f,t}$ is scaled by the generated importance coefficients via a sigmoid function $\sigma$ over the space as follows,
\begin{align}
  s_{f,t} &= \rho \hat{e}_{f,t} + \tau \\
  \hat{\vct{x}}_{f,t} &= \vct{x}_{f,t} \cdot \sigma(s_{f,t})
\end{align}
The recalibrated new feature group has all of the enhanced features. It is worth noting that the total number of $\rho$ and $\tau$ is equal to the number of groups, which is negligible compared to the millions of model parameters.

\section{Experiments}
\label{sec:exp}
\subsection{Corpora}
For this study, the VoxCeleb 1 and 2~\cite{nagrani2017voxceleb,chung2018voxceleb2} datasets are used for experiments. Models are trained on Voxceleb1 dev set and Voxceleb2 dev set, respectively.
Voxceleb1 dev is used to train the base models for ablation studies and analysis of model structure and parameters. Voxceleb2 dev is used to train our final models.
Voxceleb1 dev set contains 148,642 utterances from 1211 celebrities, and Voxceleb2 dev set has 5994 speakers with 1,092,009 utterances. Both are large scale datasets extracted from YouTube videos that are recorded across a large number of challenging visual and auditory environments, including background  conversations, laughter, overlapping speech, and a wide range of varying room acoustics. 
To evaluate the performance of the proposed global time-frequency context models for speaker representation learning, we test our models on three trial lists: (1) VoxCeleb 1-O: an original trial list containing 37,611 trials from 40 speakers in the Voxceleb 1 test set; (2) Voxceleb 1-E: an extended trial list containing 579,818 trials from 1251 speakers; and (3) Voxceleb 1-H: a hard trial list containing 550,894 trials from 1190 speakers.

\subsection{Data Augmentation}
The original data is augmented on the fly with noise and room impulse response (RIR) from the MUSAN \cite{musan2015} and OpenSLR Room Impulse Response and Noise \cite{ko2017study} corpora. There are in total 60 hours of human speech, 42 hours of music, and 6 hours of other background noise types in the MUSAN corpus. The reverberation and noise data simulation is randomly selected with equal probability. For each augmentation, one noise file is randomly selected from the MUSAN database and added to the recording with 0-15 dB SNR. Alternatively, one music file is randomly selected from the same database and added to the recording with 5-15 dB SNR. Human babble speech is selected with 13-20 dB SNR. Otherwise, a reverberation file is selected from the RIR dataset.

\subsection{Speaker Recognition Model}
\subsubsection{Model Structure}
In this study, ResNet34~\cite{heo2020clova} is used as our speaker recognition model. It is widely used in the speaker recognition community.  The residual layers’ channel sizes are 32, 64, 128, 256, respectively.
Attentive statistical pooling~\cite{okabe2018attentive} is applied to aggregate the variable-length input sequence into a fixed utterance-level speaker embedding. For the loss function, the softmax and angular prototypical loss~\cite{chung2020defence} are combined to learn discriminative embeddings.

\subsubsection{Loss Function}
\label{sec:loss}

To learn discriminative speaker embeddings, for each mini-batch, $M$ utterances are sampled from every speaker. The prototype of the speaker is defined as,
\begin{align}
\mathbf{c}_{j}=\frac{1}{M-1} \sum_{m=1}^{M-1} \mathbf{x}_{j, m}
\end{align}
During training, each example is classified against $N$ speakers based on a softmax over distances to each speaker prototype, the angular prototypical loss with a learnable scale and bias is represented as,
\begin{align}
L_{\mathrm{p}}=-\frac{1}{N} \sum_{j=1}^{N} \log \frac{e^{w \cdot cos(\mathbf{x}_{j,M}, \mathbf{c}_j) + b}}{\sum_{k=1}^{N} e^{w \cdot cos(\mathbf{x}_{j,M}, \mathbf{c}_k) + b}}
\end{align}
The number of utterances per speaker $M$ is set to 2 in our experiments. 
For the classification purpose, a multi-class softmax cross-entropy loss is used as follows,
\begin{align}
L_s = -\frac{1}{N} \sum_{i=1}^{N} \log \frac{e^{\vct{w}_{y_{i}}^{T} \mathbf{x}_{i}+b_{y_{i}}}}{\sum_{j=1}^{C} e^{\vct{w}_{j}^{T} \mathbf{x}_{i}+b_{j}}}
\end{align}
where $\vct{w}$ and $b$ are the weight and bias of the last FC layer.
The angular prototypical loss and the cross-entropy loss are combined as our speaker loss function $L = L_s + L_p$.

\subsection{Implementation Details}

It is known that system settings vary a lot for many speaker recognition solutions. Performance improvement in some studies may actually come from other modules such as data augmentation. For a fair comparison and evaluation of our proposed methods alone, we follow the data augmentation and model backbone in a well-known study \cite{heo2020clova}.
An AdamW \cite{loshchilov2018decoupled} optimizer is used with $5e^{-5}$ weight decay. A linear warm-up strategy is applied for the first 5 epochs to the initial learning rate $10^{-3}$. The learning rate is reduced by 0.75 every 18 epochs. 
The extracted utterance-level embeddings are L2 normalized, and the dimension size is $512$. 
The general dot-product scoring function is applied to compute the time-frequency attention matrix. The proposed global context model block is inserted after the last Batch Norm layer in each residual basic block. Also, cosine distance scoring is applied to evaluate verification performance. We sample 2 utterances per speaker for each mini-batch, and 500 utterances per speaker for each epoch.

We compute 64 dimensional log-mel filter-bank energies (fbank) at the frame level as input features. A Hamming window of length 25 ms with 10 ms frame shift is used to extract fbanks from input audio signals.
A chunk of 200 frames of features of each audio file is used as input to the network.
The input features are mean normalized at the frame level.

\subsection{Evaluation Metric}
Performance is reported in terms of Equal Error Rate (EER), where the false accept rate and false reject rate are equal, as well as the minimum Detection Cost Function (minDCF) with settings $P_{target} = 0.01, C_{miss} = 1, C_{fa} =1$.

\section{Results and Discussions}
\label{sec:result}

\subsection{Experimental Results}

In order to thoroughly evaluate our proposed methods, we compare all proposed models and later show a detailed ablation analysis.
The experiments are first performed on the attention based global context model (Att-GCM), followed by the group-wise time-frequency enhancement (Att-GCM-TFE), and then the DCT based global context model (DCT-GCM) for speaker verification. 
Finally, we show the analysis of various factors that may affect the performance of the proposed global context models.

A strong backbone ResNet34 model is used with ASP pooling, along with the loss function explained in Sec. \ref{sec:loss}. This achieves 2.45\% EER and 0.3471 minDCF using only the Voxceleb1 dev set. The baseline Squeeze and Excitation module can recalibrate each channel of the speech feature map to emphasize important channels and diminish insignificant ones. The computational cost of the SE module is on the order of $O(C^2/r)$, where $C$ is the number of channels and $r$ (16 in our experiments) is the reduction ratio used to reduce the model parameters. As shown in Table \ref{tab:vox1_results}, the SE module brings 3.27\% relative improvement in EER. This indicates that context modeling (GAP) and channel interactions are helpful to learn a good speaker embedding.

Table \ref{tab:vox1_results}  shows all results for our proposed models with the best settings. For the attention based context model, the proposed channel wise attention (Att-GCM) improves SV performance by a large margin compared with the ResNet34 model. With the Att-GCM block, the overall EER of the ResNet34 model decreases from 2.45\% to 2.16\%, relatively 11.84\%. Also, a relative reduction of 8.86\% EER is achieved compared with the SE block. These results may suggest that our proposed attention based global time-frequency context block can greatly recalibrate the significant feature regions to improve speaker verification performance with the context information.

\begin{table}[htbp]
  \centering
   \caption{SV results on the VoxCeleb1-O test set using our proposed attention and DCT based global context model (GCM) and time-frequency enhancenment (TFE) methods. Models are trained on the Voxceleb1 split.}
     \begin{adjustbox}{max width=1\linewidth}
    \setlength{\tabcolsep}{4pt}
    \begin{tabular}{l|ccc}
        \toprule

       Model   & \multicolumn{1}{c}{EER (\%)} & \multicolumn{1}{c}{minDCF}   \\
    \hline

    Ivector~\cite{nagrani2017voxceleb} & 8.80 & 0.7300 &  \\
    Xvector~\cite{okabe2018attentive} & 3.85 & 0.4060 &  \\
    SPE~\cite{jung2019spatial} & 4.20 & 0.4220 &  \\
    LDE~\cite{cai2018exploring} & 4.56 & 0.4410 &  \\
    MHSA~\cite{india2019self} & 4.00  & 0.4500  &   \\
    ResNet34       & 2.45  & 0.3471 &  \\	
    \hline
    ResNet34 + SE  & 2.37 &  0.3570  &  \\	
    ResNet34 + Att-GCM (ours)  & 2.16 & 0.2768 &   \\
    ResNet34 + Att-GCM + TFE (ours)  & 2.10 & 0.2602   &  \\
    
    ResNet34 + DCT-GCM  (ours) & 2.13 & 0.2503  &  \\			
    ResNet34 + DCT-GCM + TFE (ours) & 2.07 &  0.2552 &   \\																			
    \bottomrule
    \end{tabular}%
    \end{adjustbox}
  \label{tab:vox1_results}%
\end{table}%

Consistent performance improvement is also observed with a time-frequency enhancement step (Att-GCM-TFE). The EER reduces from 2.16\% to 2.10\%, with an additional relative 2.78\% improvement. It is suggested that with a good intermediate global context representation, it is possible to use the interaction between local information at each time-frequency location and global context information to adjust the speech representations. By emphasizing important time-frequency features and ignoring irrelevant features, this solution obtains more meaningful information for SID compared with calibrating channel-wise features alone.

For the DCT based context model, it is observed that 2D-DCT coefficients specific to a time-frequency location are informative for context modeling. The number of additional parameters introduced with this module is 0 since all DCT coefficients are pre-computed. 
Here, fixed length 200 frames of 64-dim fbank are used as input features, and the 2D-DCT space is divided into $8\times 25$ parts since the smallest feature map size after the last block is $8\times 25$.  

With this method, there is a relative reduction of 10.13\% and 12.66\% EER when comparing SE with DCT-GCM and DCT-GCM-TFE, respectively. This indicates that GAP does not reflect sufficient information from each time-frequency location, while the results of DCT context models are as effective as the attention context models. Another advantage is that the weights are pre-computed. Based on the observations, further detailed analysis will be presented in Sec. \ref{sec:dct_analysis}.

\subsection{Analysis of Channel Transformation}
As discussed in Sec. \ref{sec:local_channel}, for the proposed Att-GCM model, two channel transformation modules are first compared in Table \ref{tab:channel_transform}. The  ECA module only has $k$ (kernel size of 1D Conv) extra parameters, and is an effective option if a compact model is needed for on-device applications. The two FC layers brings extra $2\times C^2/r$ parameters. FC layers first projects channel features into a lower $r$-dimensional space and then maps them back. It makes the correspondence between channels and layer weights indirect. We observe that FC layers achieve the best result. In all subsequent experiments, two FC layers are used as the channel transformation with $r=16$.

\begin{table}[htbp]
  \centering
  \caption{SV results on the Voxceleb1 test set using different channel transformations of the Att-GCM block. Models are trained on the Voxceleb1 split.}
  \begin{adjustbox}{max width=1\linewidth}
    \setlength{\tabcolsep}{12pt}
  \begin{tabular}{l|ccc}
    \toprule
    Channel transform & \multicolumn{1}{c}{EER(\%)} & \multicolumn{1}{c}{minDCF} \\
    \hline
    FC layer    & 2.16 & 0.2768 \\
    ECA     & 2.24 & 0.3044 \\
    \bottomrule
    \end{tabular}
  \end{adjustbox}
\label{tab:channel_transform}
\end{table}

\subsection{Analysis of Attention Context Model}

In this section, further analysis is presented regarding factors that may affect performance of the attention context model with the TFE block.

\noindent \textbf{Normalization components} $\rho$ and $\tau$.
As shown in Table \ref{tab:norm_param}, the initialization of normalization parameters $\rho$ and $\tau$ in the 
Att-GCM-TFE block affects the verification results. 
Initializing $\rho$ to 0 tends to give better results. With a grid search, it is determined that the best setting is to assign $\rho$ to 0 and $\tau$ to 1. 
This suggests that it is appropriate to discard the context-guided time-frequency enhancement in the very early stage of network training. 
It is meaningful to first learn a good representation with the convolution stem.

\begin{table}[htbp]
  \centering
  \caption{SV results on the Voxceleb1 test set using different normalization parameters of the Att-GCM-TFE method. Models are trained on the Voxceleb1 split.}
  \begin{adjustbox}{max width=1.0\linewidth}
    \setlength{\tabcolsep}{14pt}
  \begin{tabular}{l|ccc}
    \toprule
    ($\rho$, $\tau$) & \multicolumn{1}{c}{EER (\%)} & \multicolumn{1}{c}{minDCF}  \\
    \hline
    (0, 0) & 2.13 & 0.2725   \\
    (0, 1) & \textbf{2.10} & \textbf{0.2602} \\
    (1, 0) & 2.26 & 0.2867 \\
    (1, 1) & 2.32 &  0.3110   \\
    \bottomrule
    \end{tabular}
  \end{adjustbox}

  \label{tab:norm_param}%
\end{table}%

\noindent \textbf{Group number.} We further investigate the number of groups in the Att-GCM-TFE module in Table \ref{tab:group_number}.
A limited number of groups may cause the diversity of feature representations to be limited.
We obtain the best EER and DCF values using the group number $8$. Too many groups result in a dimension reduction in the feature space, which may cause a weaker representation for each group response. The group number is set to $8$ in all our experiments.

\begin{table}[htbp]
  \centering
  \caption{SV results on the Voxceleb1 test set using different group numbers of the Att-GCM-TFE method. Models are trained on the Voxceleb1 split.}
  \begin{adjustbox}{max width=1.0\linewidth}
    \setlength{\tabcolsep}{10pt}
  \begin{tabular}{l|ccc}
    \toprule
    Group number & \multicolumn{1}{c}{EER (\%)} & \multicolumn{1}{c}{minDCF}  \\
    \hline
    4     &  2.34 & 0.3003  \\
    8     & \textbf{2.10} & \textbf{0.2602}  \\
    16    & 2.16  & 0.2699  \\
    \bottomrule 
    \end{tabular}
  \end{adjustbox}
  \label{tab:group_number}%
\end{table}%

\noindent \textbf{Block position.}
As shown in Table \ref{tab:position}, inserting the proposed module after/before the Batch Norm layer, or before the convolution layer in the residual basic block (two layers of 3x3 convolution and a residual connection) all improves the results, compared to the baseline ResNet34 and SE model. Here, only one proposed block is inserted after the Batch Norm layer in our experiments.
The Att-GCM-TFE block only requires about 0.40M additional parameters, and therefore is very computationally efficient.

\begin{table}[htbp]
  \centering
  \caption{SV results on the Voxceleb1 test set using different block positions of the Att-GCM-TFE method. Models are trained on the Voxceleb1 split.}
  \begin{adjustbox}{max width=1.0\linewidth}
    \setlength{\tabcolsep}{10pt}
  \begin{tabular}{l|ccc}
    \toprule
    Block position & \multicolumn{1}{c}{EER (\%)} & \multicolumn{1}{c}{minDCF}  \\
    \hline
    after BN & \textbf{2.10} & \textbf{0.2602}   \\
    before BN & 2.29	& 0.2865   \\
    before Conv & 2.32 & 0.3183    \\
    \bottomrule
    \end{tabular}
  \end{adjustbox}

  \label{tab:position}%
\end{table}%

\begin{figure}[htbp]
    \centering
    \includegraphics[width=0.9\linewidth, height=5.5cm]{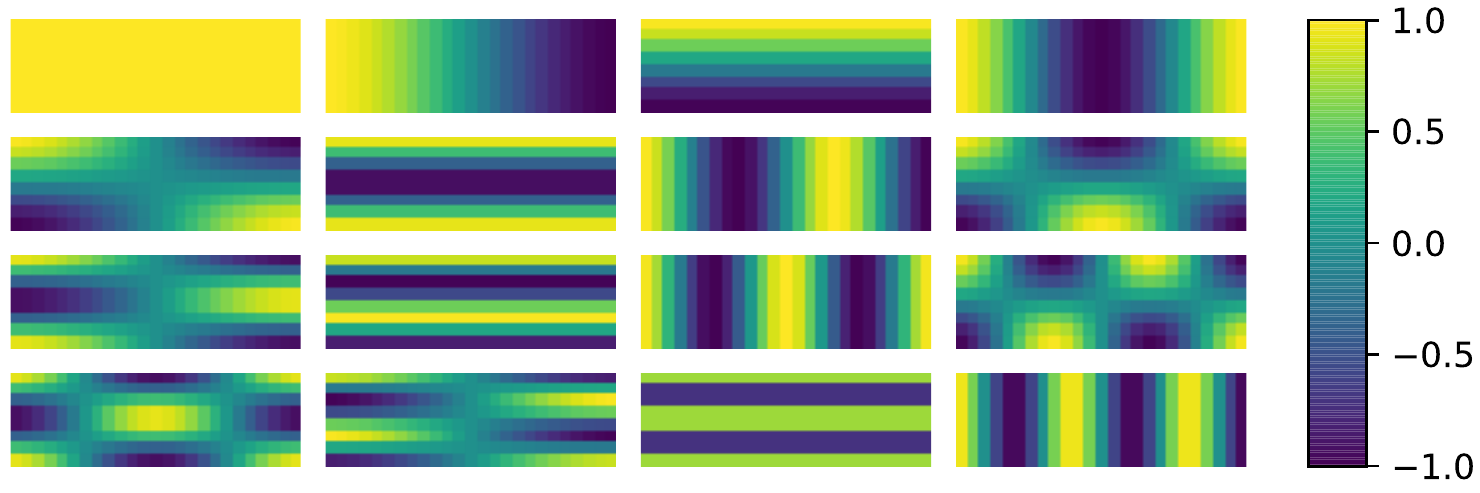}
    \caption{Visualization of the lowest 16 DCT components of the $8 \times 25$ DCT bases, order from 1st lowest to the 16th lowest along the row dimension. The first two are the two lowest DCT components used in our models.}
    \label{fig:selec_dct}
\end{figure}

\subsection{Analysis of DCT Context Model}
\label{sec:dct_analysis}

As noted in Sec. \ref{sec:multi_dct}, the $K$ lowest DCT components are chosen as the bases.
We compare [1,2,4,8,16,32] lowest frequency components and show the result in Table~\ref{tab:dct-freq}. 
1 DCT basis corresponds to the original GAP in the channel attention. In this case, all other DCT components are discarded. It is noted that with  frequency components and a maximum activation function, it is possible to embed more time-frequency information into one channel. When using 2 DCT components, our model obtains the best result with a 2.13\% EER and 0.25 DCF, which is relatively 10.13\% and 29.89\% lower than GAP.
Also, when two channels are redundant in the speech representation, we may obtain the same information using GAP. However, in the proposed multi-DCT attention method, we can get more information since different DCT components can be selected. 
Weights of each time-frequency location are computed from the DCT base functions. Discrete Cosine Transform corresponds to a signal compression on the 2D feature map. With the pre-computed DCT coefficients, we know how the 2D feature matrix is represented by DCT values. This is helpful for us to probe the model compared with using the attention mechanism as a black box. DCT isolates the most relevant features and patterns within the data, effectively de-emphasizing noise and redundancies. This, in turn, enables a more accurate and efficient speaker recognition system.

\begin{table}[htbp]
  \centering
   \caption{SV results on the VoxCeleb1 test set using the DCT-GCM method with different numbers of DCT components. Models are trained on the Voxceleb1 split.}
    \begin{adjustbox}{max width=1.2\linewidth}
    \setlength{\tabcolsep}{14pt}
    \begin{tabular}{l|ccc}
    \toprule
    Number   & \multicolumn{1}{c}{EER (\%)} & \multicolumn{1}{c}{minDCF} \\
    \hline
    1 & 2.37 &  0.3570  \\
    2 & \textbf{2.13} & \textbf{0.2503}    \\
    4 & 2.27 &  0.3110    \\			
    8 & 2.23 & 0.2813     \\
    16 & 2.23  & 0.2914    \\
    32 & 2.32  & 0.3236     \\
    \bottomrule
    \end{tabular}%
    \end{adjustbox}
  \label{tab:dct-freq}%
\end{table}%

\begin{figure*}[t]
\centering
\subfigure[4s ${\textcolor[RGB]{255,54,160}{\blacksquare}}$]{\includegraphics[width=0.445\linewidth, height=7cm]{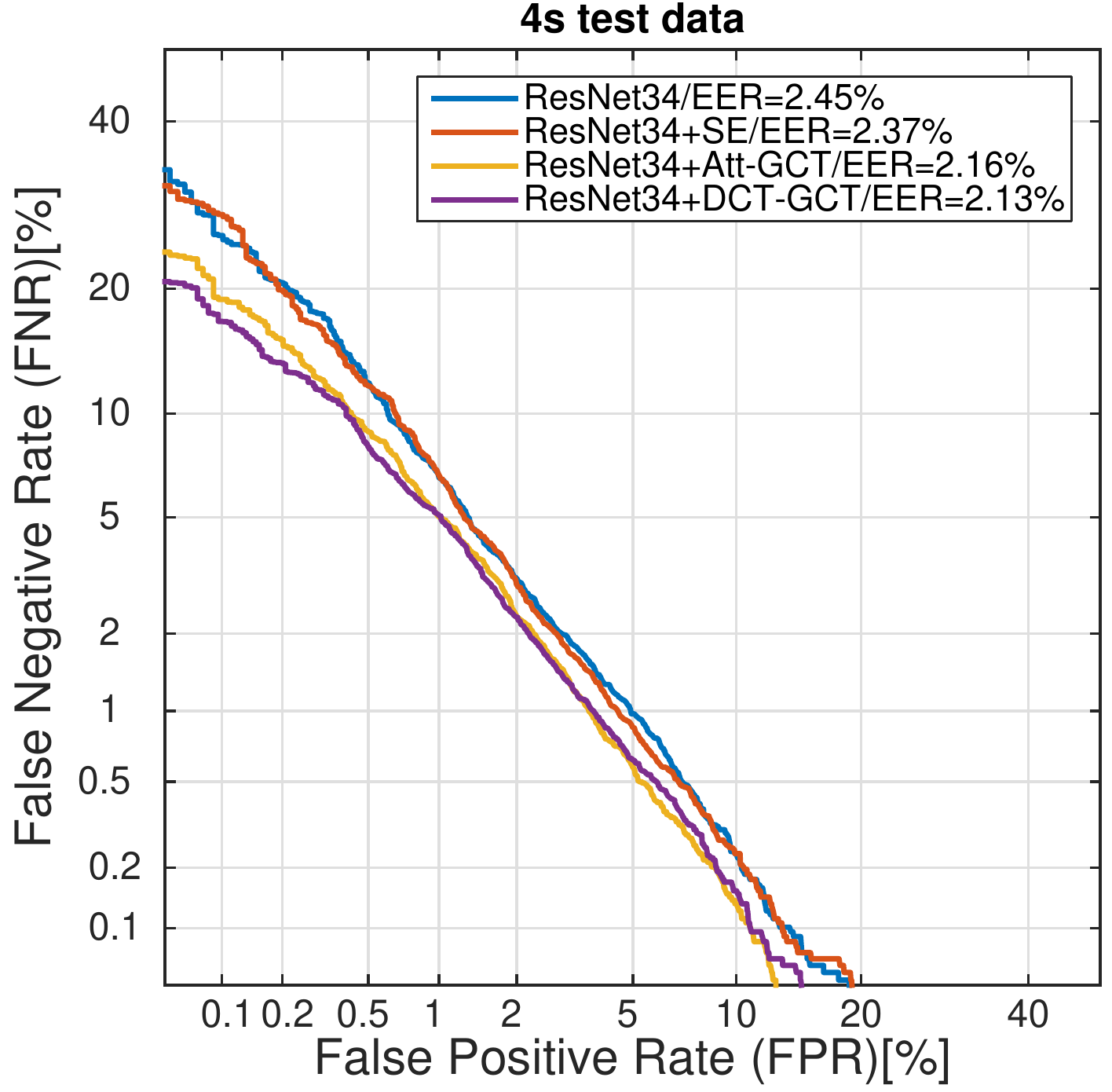}}
\subfigure[3s ${\textcolor[RGB]{96,17,0}{\blacksquare}}$]{\includegraphics[width=0.445\linewidth, height=7cm]{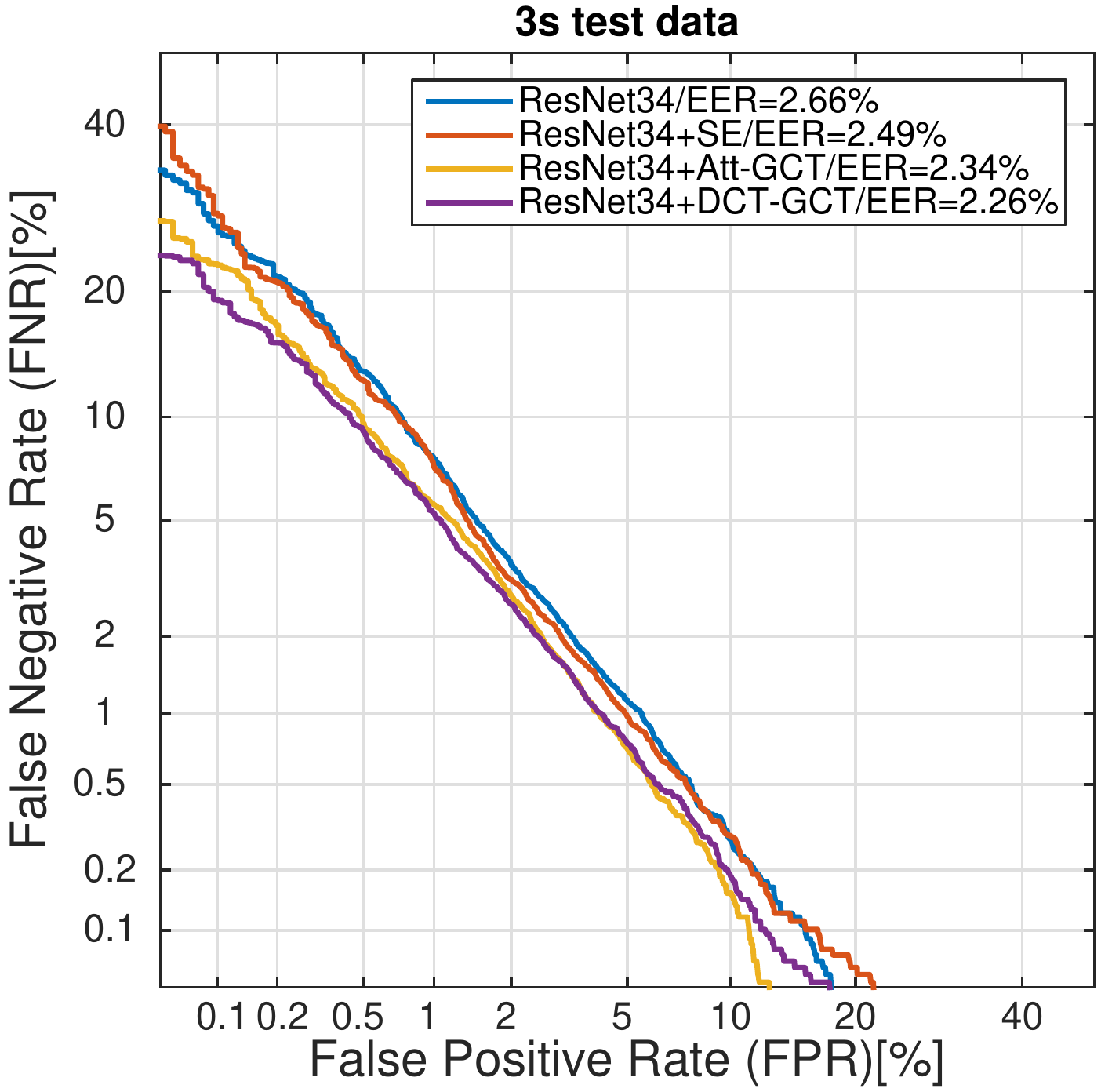}}
\subfigure[2s ${\textcolor[RGB]{255,125,10}{\blacksquare}}$]{\includegraphics[width=0.445\linewidth, height=7cm]{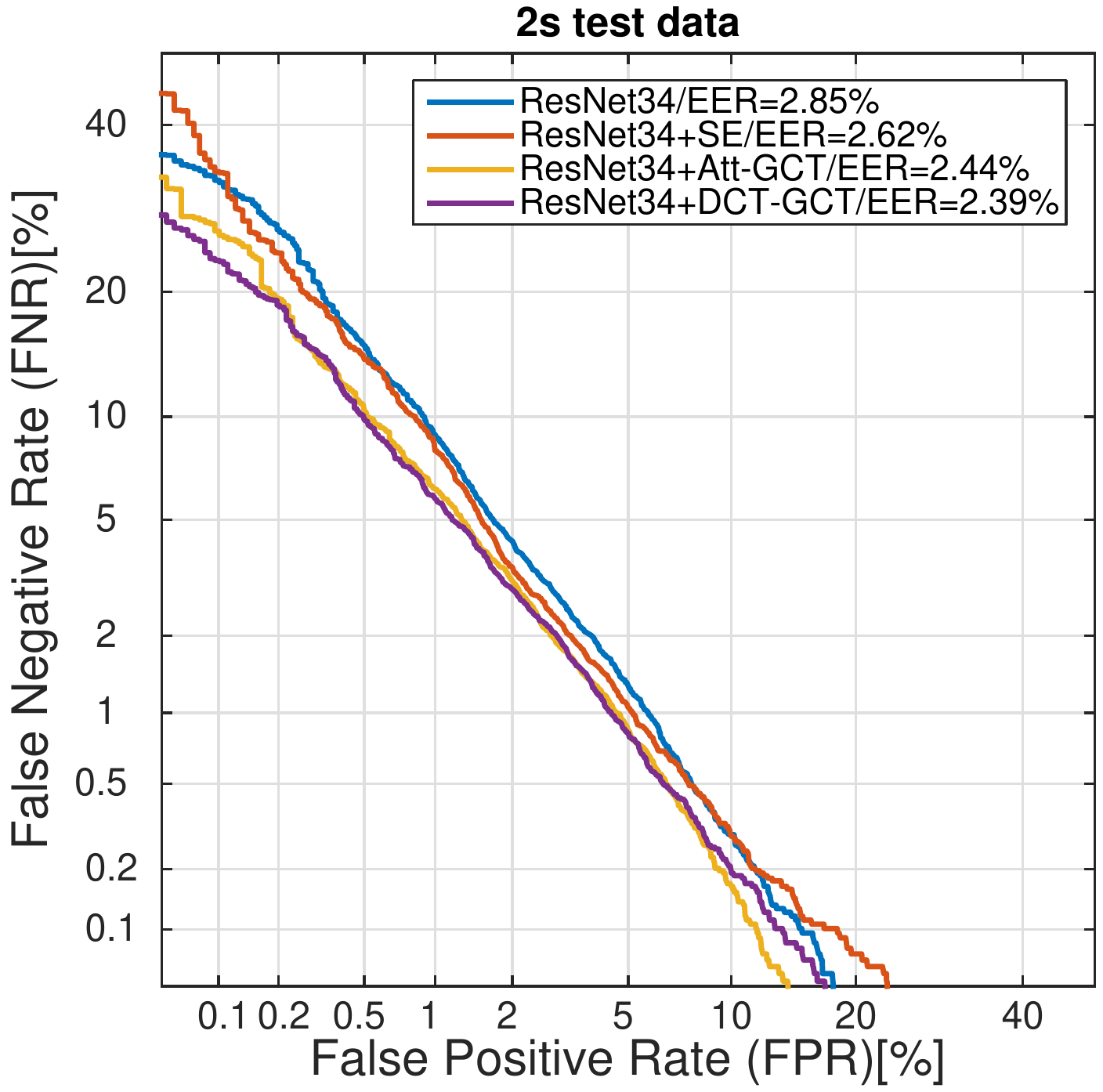}}
\subfigure[1s ${\textcolor[HTML]{3399FF}{\blacksquare}}$]{\includegraphics[width=0.445\linewidth, height=7cm]{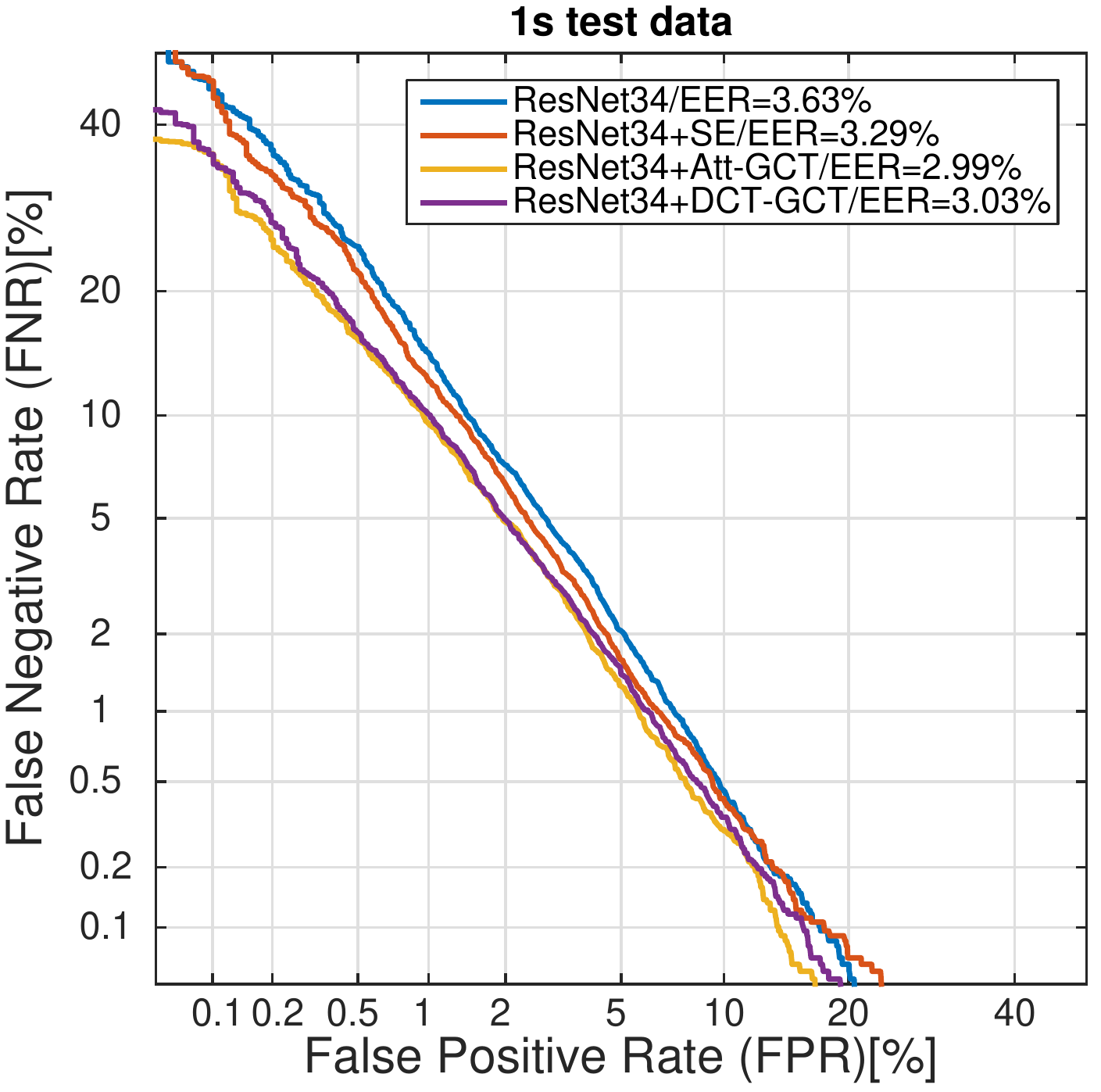}}

\captionsetup{font={footnotesize}}
\caption{Detection Error Trade-off (DET) curves for each system on the Voxceleb 1 test set with different test data duration - 4, 3, 2, 1s respectively.}
\vspace{1.8ex}
\label{fig:det_duration} 
\end{figure*}

\subsection{Visualization of Selected DCT Components}
In Fig. \ref{fig:selec_dct}, we show the 16 lowest DCT base functions of $8 \times 25$ DCT parts. These are ordered from the 1st to the 16th lowest along the row. The figure shows that 2D DCT basis functions are composed of regular horizontal and vertical cosine wave functions. These DCT components are orthogonal and data-independent. The DCT based global context model improves model efficiency and explainability. Compared with the attention based global context model, all DCT weights are pre-computed. Therefore, the overall model complexity is significantly reduced. The first two components in the first row are the two lowest components, which are used in our experiments for DCT based context models.

\begin{table*}[tbp]
  \centering
  \caption{SV results on the VoxCeleb1 test set using  models with our proposed global time frequency context blocks. Speaker models are trained on the Voxceleb2 dev split.}
    \begin{adjustbox}{max width=0.95\linewidth}
    \setlength{\tabcolsep}{10pt}

    \begin{tabular}{lcccccccl}
    \toprule
    \multirow{2}[4]{*}{Model} & \multicolumn{2}{c}{VoxCeleb1-O} & \multicolumn{2}{c}{VoxCeleb1-E} & \multicolumn{2}{c}{VoxCeleb1-H} &  \\
\cmidrule{2-7}          & \multicolumn{1}{l}{EER (\%)} & \multicolumn{1}{l}{minDCF} & \multicolumn{1}{l}{EER (\%)} & \multicolumn{1}{l}{minDCF} & \multicolumn{1}{l}{EER (\%)} & \multicolumn{1}{l}{minDCF} &  \\
    \midrule
 
    E-TDNN \cite{snyder2019speaker}  &  1.49  &   0.1604   &   1.61   & 0.1712 &   2.69    &  0.2419  &  \\
    ARET-25 \cite{zhang2020aret} &  1.39  &    N/A   &   1.52    & N/A &   2.61    &  N/A  &  \\
    ECAPA-TDNN (C=512)  &  1.08  &    0.1251   &  1.25    & 0.1574 &   2.46    & 0.2601  &  \\

    \hline
    ResNet34 + SE  & 1.10  &  0.1254  & 1.22 & 0.1556 &  2.40 & 0.2593 &  \\

    ResNet34 + Att-GCM (ours)  &  0.98   & 0.1186  & 1.14 & 0.1452 & 2.26  &  0.2482 &  \\ 
    ResNet34 + Att-GCM + TFE (ours) & 0.94  & 0.1075   & 1.10 &  0.1406   & 2.25 & 0.2407 &  \\
    ResNet34 + DCT-GCM (ours)  & 0.97  &  0.1143 &  1.12 &  0.1430  & 2.28 & 0.2494 &  \\
    ResNet34 + DCT-GCM + TFE (ours)  &  0.90 &  0.1036  & 1.08 &  0.1392  & 2.22 & 0.2410      &  \\

    \bottomrule
    \end{tabular}%
    \end{adjustbox}
  \label{tab:vox2}%
\end{table*}%

\subsection{Duration Analysis}

Next, in Fig. \ref{fig:det_duration}, performance of the four systems is shown for a range of test data durations. In general, performance degrades when the test data duration becomes smaller. It is especially challenging to maintain effective performance in very short test conditions. When the duration is very short (e.g., 1s), context models and SE are all vulnerable. In terms of EER, attention and DCT based global context models are more robust for short duration data than the SE model. From ResNet34 to ResNet34+SE system, there is an average of 6.77\% relative improvement in EER across different durations.  The Att-GCM and DCT-GCM context models obtain additional 13.97\% and 15.19\% improvement on average, respectively. Also, a high false rejection rate means that the same speakers are incorrectly classified since they are very near in the cosine distance space. This suggests that if it is possible to further improve  discriminative ability in the speaker embedding (e.g., by introducing more utterances per speaker or adding more augmentation), there is still room to improve performance. Deduced from Fig. \ref{fig:det_duration}, in most occasions, speaker recognition systems are vulnerable when test data are too short. It would be better to filter out very short data when collecting data in real scenarios. It also suggests that our speaker model might learn more meaningful information with global context modules and reduce the impact of duration mismatch. The proposed systems might be helpful in some very short duration speaker or keyword detection applications.

\subsection{Results with Models Trained on Voxceleb2 Data}
Based on comprehensive analysis and ablation studies in previous sections, we show the results of our best proposed models trained on the Voxceleb2 dev split, and compare with other public results, including E-TDNN \cite{snyder2019speaker,desplanques2020ecapa}, ARET \cite{zhang2020aret}, and ECAPA-TDNN \cite{desplanques2020ecapa} (re-implemented following the original paper).
From Table \ref{tab:vox2}, it can be observed that a significant improvement with our proposed attention and DCT based context models is achieved. Compared with the baseline SE block, there is an average of 10.21\% relative improvement in EER over three test trials using the attention context model Att-GCM-TFE, and 12.38\% relative improvement using the DCT context model DCT-GCM-TFE. Results of DCT based global context models are as good as attention based context models. The difference is that DCT coefficients are pre-computed with no increase in computational costs. With these two convenient modules, our approach can easily improve speaker recognition performance by emphasizing important regions. 
The attention context model is purely data driven, so it may obtain better performance in some conditions by emphasizing salient regions. It does not require additional DCT component selection or a mapping between the feature map and the DCT indices.

\section{Conclusions}
\label{sec:conclusion}
In this study, a global time-frequency context modeling method has been proposed and successfully applied to both the channel and time-frequency wise feature map recalibration. Our proposed global context models mainly include  attention-based and DCT based context models. A time-frequency enhancement method was also proposed to leverage the correlation between global context and local feature vectors at each time-frequency to guide the speaker representation learning better. The attention context model can capture long-range time-frequency dependency and channel variances. This lightweight block was shown to enhance the latent speaker representation and suppress possible distortions. The block was inserted after the last Batch Norm layer of each residual basic block. The DCT context model used the pre-computed DCT coefficients as weights. A multi-DCT approach was proposed to use different DCT components. The solution also illustrated that GAP in SE is a special case of the formulated attention and DCT based context models. Experimental results were significantly improved by emphasizing the import channel and time-frequency regions with our proposed methods.  

Extensive experiments, ablation studies, and analysis were performed to evaluate the proposed methods on Voxceleb data. The approaches were shown to improve the ResNet and SE models in terms of both EER and DCF by a large margin. We found that using the FC layer for channel transformation obtained the best result compared with 1D Conv. For on-device applications, choosing 1D Conv with a small kernel size $K$ would be worthwhile since it can reduce the additional parameters introduced by channel transform from an order of $O(C^2/r)$ to $O(K)$. The time-frequency enhancement method can further improve the proposed context models by strengthening salient time-frequency locations. We also discussed the impact of the normalization components, group number, and block position on the Att-GCM-TFE method. 
For our proposed DCT context models, we found that we could get more information with multiple DCT components. It is suggested to use the two lowest frequency components with the proposed method. The proposed global context models not only achieve better performance compared with the SE block, but are also generally more robust to short duration test data in general. They might also be helpful for short duration speaker or keyword detection applications.

The proposed global context models can effectively recalibrate the feature maps adaptively by emphasizing more important channels and time-frequency locations. For future work, we will consider leveraging second-order pooling and pooing over time approaches for the feature recalibration. At the same time, this approach might be applied to many other applications, such as language identification, speaker adaption for speech recognition, spoofing detection, and emotion recognition. Therefore, this study highlights effective global context methods for text-independent speaker recognition and fundamental observations for future studies.

\bibliographystyle{IEEEtran}
\bibliography{main.bib}

\end{document}